\documentstyle[12pt]{article}
\begin{document}
\title{ON THE AREA DISTANCE AND THE
 RIEMANNIAN GEOMETRY}
\author{D. PALLE\thanks{ 
Zavod za teorijsku fiziku, Institut Rugjer Bo\v {s}kovi\'{c}, 
P.O.Box 1016, Zagreb, Croatia}}  
\date{ }
\maketitle
 We study an area distance in the Riemannian
 spacetime with expansion, vorticity and acceleration.
It is shown that this observable depends on expansion, deceleration
 and acceleration parameters to third order in redshift, as well as
 on vorticity. Thus, the considerable acceleration 
 or vorticity of the Universe can
 substantially influence the standard astronomical distance calibrators. \\
\\
\begin{eqnarray*}
Cosmology-Riemannian\ geometry-Area\ distance.
\end{eqnarray*}

\vspace{6 mm}
In the standard cosmological model, one usually deals with the Robertson-Walker
line element to describe various kinematical, dynamical or alchemical
 processes in the Universe. Although the observations refer to the high
degree of homogeneity and isotropy of the Universe, current and future
measurements would require theoretical models with a more general
geometry of spacetime. A fundamental role in all kinds of astronomical
observation plays area, luminosity, angular diameter or comoving distances.
In this paper we choose an area distance to investigate
its dependence on the parameters of the Riemannian geometry.

We start with the definition of the area(corrected luminosity) distance due
to Kristian and Sachs \cite{Krist} :

\begin{eqnarray}
r^{2}&=&\frac{I_{z}}{I_{0}}(\frac{f_{0}}{f_{z}})^{4},\hspace*{45 mm} \\
r&=&corrected\ luminosity\ distance, \nonumber \\
I_{(0;z)}&=&(measured;intrinsic)\ intensity, \nonumber \\
f_{(0;z)}&=&(measured;intrinsic)\ frequency, \nonumber \\
D^{2}&=&(1+z)^{4}r^{2}, \nonumber \\
D&=&luminosity\ distance. \nonumber
\end{eqnarray}

For the quotient of intrinsic and measured frequencies, the following
relation is valid:

\begin{eqnarray}
\frac{f_{z}}{f_{0}}=\frac{(u_{\mu}k^{\mu})_{z}}{(u_{\mu}k^{\mu})_{0}}, 
 \hspace*{60 mm} \\
u^{\mu}=four-velocity\ vector,
\ k^{\mu}=vector\ tangent\ to\ geodesics. \nonumber 
\end{eqnarray}  

Expanding various quantities with respect to the affine parameter, 
Kristian and Sachs derived the formula that relates the corrected
luminosity(area) distance and redshift \cite{Krist}:

\begin{eqnarray}
\frac{f_{z}}{f_{0}}&=&1-(u_{\mu;\nu})_{0}e^{\mu}e^{\nu}r-
\frac{1}{2}(u_{\mu;\nu;\kappa})_{0}e^{\mu}e^{\nu}e^{\kappa}r^{2} \nonumber  \\
&-&(\frac{1}{6}(u_{\mu;\nu;\kappa;\rho})_{0}e^{\mu}e^{\nu}e^{\kappa}e^{\rho}
+\frac{1}{12}(u_{\mu;\nu})_{0}e^{\mu}e^{\nu}(R_{\kappa\rho})_{0}
e^{\kappa}e^{\rho})r^{3}+ ... \\
\frac{f_{z}}{f_{0}}&\equiv &1+z,\ u_{\mu;\nu}\equiv\partial _{\nu}u_{\mu}-
\Gamma^{\kappa}_{\mu\nu}u_{\kappa}, \nonumber \\
e^{\mu}&\equiv &-(k^{\mu}/u_{\nu}k^{\nu})_{0},\ R_{\mu\nu}=Ricci\ tensor. \nonumber
\end{eqnarray}

We use this formula with the coefficients that are averaged over the
directions of tangents to geodesics

\begin{eqnarray}
z&=&-\langle (u_{\mu;\nu})_{0}e^{\mu}e^{\nu}\rangle r
-\frac{1}{2}\langle (u_{\mu;\nu;\kappa})_{0}e^{\mu}e^{\nu}
e^{\kappa}\rangle r^{2} \nonumber \\ &-&\langle
\frac{1}{6}(u_{\mu;\nu;\kappa;\rho})_{0}e^{\mu}e^{\nu}e^{\kappa}e^{\rho}
+\frac{1}{12}(u_{\mu;\nu})_{0}e^{\mu}e^{\nu}(R_{\kappa\rho})_{0}
e^{\kappa}e^{\rho}\rangle r^{3}+ ... \\
\langle F(\theta,\phi)\rangle& \equiv &\frac{1}{4\pi}\int F(\theta,\phi)d\Omega,
\nonumber \\
u^{\mu}&=&(1,0,0,0),\ k^{\hat{\mu}}=(1,sin\theta cos\phi,sin\theta sin\phi,
cos\theta), \nonumber \\
k^{\mu}&=&h^{\mu}_{\cdot\hat{\nu}}k^{\hat{\nu}},\ 
h^{\mu}_{\cdot\hat{\nu}}=vierbein.  \nonumber
\end{eqnarray}

Let us now choose the shearless metric with nonvanishing expansion,
 vorticity and acceleration \cite{Korot}:

\begin{eqnarray}
ds^{2}=dt^{2}-2\sqrt{l}R(t)e^{mx}dtdy-R^{2}(t)(dx^{2}+
b e^{2mx}dy^{2}+dz^{2}), \\
b,l,m\ are\ constant\ parameters. \hspace*{20 mm} \nonumber
\end{eqnarray}

The standard form of the covariant derivative of the four-velocity
vector looks like \cite{Ehle}:

\begin{eqnarray}
u_{\mu;\nu}&=&\frac{1}{3}\Theta h_{\mu\nu}+\omega_{\mu\nu}+
\sigma_{\mu\nu}+\dot{u}_{\mu}u_{\nu},  \hspace*{30 mm} \\
\dot{u}_{\mu}&\equiv &u^{\nu}u_{\mu;\nu},\ \Theta\equiv u^{\nu}_{ ;\nu},
\ h_{\mu\nu}=g_{\mu\nu}-u_{\mu}u_{\nu},\nonumber \\
\omega_{\mu\nu}&\equiv &\frac{1}{2}(u_{\rho;\kappa}-u_{\kappa;\rho})
h^{\rho}_{\cdot\mu}h^{\kappa}_{\cdot\nu}, \nonumber
\end{eqnarray}

 with the following observables for the chosen metric:

\begin{eqnarray}
\sigma&\equiv &\frac{1}{2}(\sigma_{\mu\nu}\sigma^{\mu\nu})^{\frac{1}{2}}=0,\ 
\omega\equiv \frac{1}{2}(\omega_{\mu\nu}\omega^{\mu\nu})^{\frac{1}{2}}=
\frac{m}{2R(t)}(\frac{l}{b+l})^{\frac{1}{2}}, \nonumber \\
\Theta&=&3\frac{\dot{R}}{R}=3 H,\ \dot{u}^{\nu}\dot{u}_{\nu}=-\Sigma H^{2},\ 
\Sigma\equiv \frac{l}{b+l}. \nonumber
\end{eqnarray}

Performing derivatives, summations and angle averaging we get  
the coefficients of Eq.(4). Inverting Eq.(4) one can immediately
find an area distance as power series in redshift:

\begin{eqnarray}
H_{0}r&=&z-\frac{1}{2}(3+q_{0})(1+\frac{\Sigma}{3})z^{2}+ 
[\frac{1}{2}(1+\frac{\Sigma}{3})^{2}(3+q_{0})^{2} \nonumber \\
&-&\frac{1}{6}(16+c_{0}+11q_{0})(1+\Sigma)+
(-\frac{1}{2}+\frac{2}{9}\frac{1}{\Sigma})\frac{\omega_{0}^{2}}
{H_{0}^{2}}]z^{3}+ ... \\
q_{0}&\equiv &-(\frac{\ddot{R}R}{\dot{R}^{2}})_{0},\ 
c_{0}\equiv (\frac{\stackrel{...}{R}R^{2}}{\dot{R}^{3}})_{0}.
\nonumber
\end{eqnarray}

From Eq.(7) one can see that besides the dependence of the corrected
 luminosity distance on the 
 Hubble parameter $H_{0}$, parameter $c_{0}$ and
the deceleration parameter $q_{0}$, there is also the dependence of $r$
on the acceleration 
parameter $\Sigma$ and the vorticity $\omega_{0}$.

Thus, considerable acceleration or vorticity of the Universe could drastically
affect astronomical distance calibrations and consequently change
our conclusions on further correlated  cosmological observables.
Our knowledge of the acceleration parameter is rather poor, but we now
mention the possibilities how to extract some information.

Assume that a relation between a Hubble's expansion and a vorticity
is valid, as it is derived in the Einstein-Cartan gravity
\cite{Palle}(this is the only theoretical consideration that gives a
small and nonvanishing cosmological constant):

\begin{eqnarray}
|\omega_{0}|\simeq\frac{\sqrt{3}}{2}\Sigma H_{0},\ 
 \hspace*{45 mm} \\
\rho_{0}\simeq\frac{3}{4\pi G_{N}}H^{2}_{0},\ 
\Lambda\simeq -\frac{1}{2}\rho_{0}.  \nonumber \hspace*{25 mm} 
\end{eqnarray}

On the other hand, the polarization measurements \cite{Korot} or
 the apparent periodicity of large-scale structures \cite{Korot2}
 gives the estimate of the vorticity $\frac{\omega_{0}}{H_{0}} = {\cal O}
 (1)$, thus suggesting a large acceleration parameter $\Sigma =
 {\cal O}(1)$.
 Some different analyses give much smaller vorticity \cite{Birch} and
 consequently smaller $\Sigma={\cal O}(10^{-2})$, if the relation (8) is valid.

One should also notice bounds put on space-rotations\cite{Barr} in 
cosmological models with vanishing vorticity \cite{Ellis} to avoid any
confusion with our choice of spacetime.

We see that the present knowledge of the parameters of the 
Riemannian cosmology is far from being resolved. Furthermore, the Einstein
field equations in the perfect fluid model
with the expansion and acceleration give $\Sigma = \frac{1}{2}$ and 
$q_{0} = -1$.
The introduction of vorticity into field equations
requires the introduction of a cosmic magnetic field into the matter
part of the equations.

To conclude, we stress that besides the ongoing astronomical measurements
of the large-scale structure of the Universe, such as the Supernovae
redshift projects, SDSS, 2dF survey, QSO redshift surveys, etc, cosmologists
have to perform calculations on the most general cosmological models
in order 
to do an overall fit of large data-sets.
\newline
\hspace*{59mm}$\ *\ *\ *\ $
\newline
This work was supported by the Ministry of Science and Technology of
the Republic of Croatia under Contract No. 00980103.

\end{document}